\documentstyle[12pt,draft]{article}

\newcommand \be {\begin{equation}}
\newcommand \bea {\begin{eqnarray}}
\newcommand \ee {\end{equation}}
\newcommand \eea {\end{eqnarray}}
\newcommand \ch { {\cal H} }
\newcommand \cd { {\cal D} }
\newcommand \usv {\frac{1}{V}}
\newcommand \pap {\frac{1+m^\alpha_i}{2}}
\newcommand \pbp {\frac{1-m^\alpha_i}{2}}

\begin{document}

\title{Maximal Mean Field Solutions in the Random Field Ising Model:
the Pattern of the Symmetry Breaking}

\author{Marco GUAGNELLI$^{(a)}$, Enzo MARINARI$^{(a,b)}$\\
and\\
Giorgio PARISI$^{(c)}$ \\[1.5em]
$(a)$: Dipartimento di Fisica and Infn,\\
Universit\`a di Roma {\it Tor Vergata},\\
Viale della Ricerca Scientifica, 00173 Roma, Italy\\[0.5em]
$(b)$: Department of Physics and NPAC, \\
Syracuse University\\
Syracuse, NY 13244, USA\\[0.5em]
$(a)$: Dipartimento di Fisica and Infn,\\
Universit\`a di Roma {\it La Sapienza},\\
P. Aldo Moro 2, 00185 Roma, Italy\\[0.5em]
{\footnotesize \tt guagnelli@roma2.infn.it marinari@roma1.infn.it
        parisi@roma1.infn.it}\\[1.0em]}
\date{March 20, 1993}
\maketitle

\vfill
\newpage

\begin{abstract}

In this note we study the mean field equations for the $3d$ Random
Field Ising Model. We discuss the phase diagram of the model, and we
address the problem of finding if such equations admit more than one
solution.  We find two different critical values of $\beta$: one where
the magnetization takes a non-zero expectation value, and one where we
start to have more than one solution to the mean field equation. We
find that, inside a given solution, there are no divergent correlation
lengths.

\end{abstract}

\vfill

\begin{flushright}
  {\bf ROM2F-92-46}\\
  {\bf SCCS 372}\\
  {\bf cond-mat/9303042 }\\
\end{flushright}

\newpage

\section{Introduction}

The Random Field Ising Model (RFIM) (see for example refs.
\cite{BELYOU,SHAPIR,PARISA}) is waiting for pieces of new
understanding and further clarifications of the relevant physical
mechanisms.

Let us start by sketching the theoretical situation. For a certain
time it was hoped that dimensional reduction could be the appropriate
method to compute the critical behavior of a ferromagnet in presence
of a random magnetic field.  It was proven in \cite{YOUNG} that in
perturbation theory the sum of the most divergent diagrams close to
the phase transition for a random field model in dimension $D$
coincides with that of a ferromagnetic theory, without random field,
in the reduced dimension $d=D-2$. The terms that are neglected are
less singular than the leading ones by a factor $\xi^{-2}$, $\xi$
being as usual the correlation length. This result suggests that all
the exponents of the random field system coincide with those of the
corresponding ferromagnetic system in $D-2$ dimensions.

Clearly this result cannot be correct.  Simple physical arguments
(confirmed by a rigorous analysis \cite{IMBRIE}) lead to the
conclusion that the lower critical dimension is $2$, not $3$, as
implied by dimensional analysis.  The deep reason for this failure can
be found following the non-perturbative analysis of ref.
\cite{PARISA,PARSOU}. Let us summarize the main results.

We assume that the system is described by the following Hamiltonian
density, which is a functional of the order parameter $\phi(x)$:

\be
  H[\phi] = \int d^Dx \ \big (
            \frac{1}{2} (\partial \phi)^2 +V(\phi) - h(x) \phi(x) \big )\ ,
\ee
where the random field $h(x)$ is a Gaussian uncorrelated
white noise with variance $g \ \delta (x-y)$, and $g$
parametrizes the strength of the random field.

The stationary points of $H$ can be found by solving
the corresponding mean field equations

\be
  -\Delta \phi + V'(\phi)=h(x)\ . \label{EQU_MF}
\ee

When these equations admit only one solution, as it happens for
sufficiently large temperature, it is natural to introduce the
correlations functions

\bea
  \nonumber C(x)&\equiv&  \overline {\phi(x)\phi(0)}\ , \\
            G(x)&\equiv& {\overline{\delta \phi(x) \over \delta h(0)}}
   = \overline{\langle x|{1 \over \Delta + V''(\phi)}|0\rangle}\ ,
\eea
where by the long bar we denote the thermal average over all the realizations
of the random magnetic fields.

These two correlation functions are the mean field approximation to
$\overline {<\phi(x)><\phi(0)>}$ and $\overline {<\phi(x)\phi(0)>_c}$
respectively. Then one finds that $C(x)$ is proportional to
the same correlation function of the pure system
in dimensions $d=D-2$. The functions
$G(x)$ and $C(x)$ are related one to the other. In Fourier space one
finds that

\bea
  \nonumber C(k)&=& g \int d\mu \ \rho(\mu)\ {1 \over (k^2+\mu^2)^2}\\
            G(k)&=& \ \ \int d\mu \ \rho(\mu)\ \ {1 \over k^2+\mu^2}\ .
\eea

The function $G(k)$ is the same as for the pure system in dimension
$d=D-2$ (dimensional reduction works in configuration space with the
function $C$, and in momentum space with the function $G$).

When \ref{EQU_MF} admits more than one solution
to compute expectation values we must assign a weight to each
solution. This makes life more complicated.  If we label by $\alpha$ different
solutions, and by $w_\alpha$ the relative weight we can write

\bea
\nonumber C(x) &=& \sum_\alpha w_\alpha \ \overline {\phi(x)\phi(0)}\\
\nonumber G(x) &=& \sum_\alpha w_\alpha \
         \overline{{\delta\phi_\alpha(x)\over \delta h(0)}} \\
     &=& \sum_\alpha w_\alpha \
         \overline{\langle x\mid{1 \over -\Delta + V''(\phi)}\mid y\rangle}
\eea

By using different prescriptions for the
weights $w_\alpha$ we can obtain different results. This is especially true
if the number of different solutions of the mean field equations
increases with the volume.

Dimensional reduction can still hold, but with a crazy choice of the weights:

\be
  w_\alpha= \frac{ {\rm sign}\  {\rm det} [-\Delta +
                    V''(\phi_\alpha)]}{Z_w}\ ,
\ee
where $Z_w$ is such that $\sum_\alpha w_\alpha =1$.
Here the Morse theorem states that $Z_w=1$.

A physically motivated choice would be:

\be
  w_\alpha= \frac { {\rm sign}\  {\rm det} [-\Delta +
              V''(\phi_\alpha)] \exp(-\beta H[\phi_\alpha])}{Z_w},
\ee
where $\alpha$ runs over all the solutions of the mean field equations,
minima, maxima and saddle points all together. The strange looking
unusual factor ${\rm sign} \ {\rm det} [-\Delta + V''(\phi_\alpha)]$
is needed to
keep the continuity of $Z_w$ when new solutions appear.

It could be
argued that the energy of the minima is so smaller than the energy
of the saddle points and maxima, that we can simply write

\be
  w_\alpha= \frac{\exp (-\beta H[\phi_\alpha])}{Z_w},
\ee
and keep the sum restricted only to the minima. In the rest of this
paper we follow this second strategy.

It is possible that this modified mean field theory gives the
correct results (as it is implicit in the work of ref. \cite{OGIELS}) and
that the failure of dimensional reduction is simply related to the
existence of many solutions with different energy
(\cite{PARISA,PARSOU,DOTPAR}).

Our aim is here to investigate numerically this improved mean field
approximation to make its predictions explicit and eventually
to compare them with Monte Carlo simulations. We have been motivated
to start this investigation by an interesting paper \cite{MEZYOU}, in
which it was suggested that replica symmetry is already broken at the
point ferromagnetic phase transition. For results obtained both in the
mean field framework and with a Monte Carlo and a $T=0$ optimization
approach, see refs. \cite{MEANFI,YOUNAU,OGIELS}.

In this note we limit ourselves to the study of two particular
solutions of the mean fields equations, which we call $\phi_+$ and
$\phi_-$.  They are such that for any solution $\phi_\alpha$ (and for
any $x$) the relation $\phi_-(x) \le \phi_\alpha(x) \le \phi_+(x)$
holds.  The existence of two solutions with this property (in the high
temperature phase they coincide) follows from convexity
arguments~\cite{AIZ}.  We call them {\em maximal mean field
solutions}.

\section{Lattice Mean Field Equations}

We consider the Random Field Ising Model (RFIM) with Ising type
($Z_2$) variables defined on a $3d$ simple cubic lattice.  We study
the solutions of its mean field equations.

 With $i$ we denote the
triplet of integers $(x,y,z)$, which characterize the lattice sites.
We will consider configurations of the random field $\{h_i\ \equiv
\theta_i\ \ch \}$, where the quenched random variables $\theta_i$ can
take the values $\pm 1$ with probability $\frac{1}{2}$, and we have
chosen the absolute value of the field, $\ch$, to be $1.5$.  Such
choice for $\ch$ was meant to allow the critical temperature $T_c$ to
have a non negligible shift from $T_c$ in the pure model, and
simultaneously not to be large enough to allow the transition
to become first order$^{\cite{BELYOU}}$.

In the mean field approximation one introduces local magnetization
variables $m_i$, which play the same role of $\phi(x)$ in the
continuum formalism.  The total free energy is written as a function
of the local magnetization, and the condition for the
free energy being stationary is the usual mean field equation

\be
  m_i = \tanh(\beta(\cd m_i + h_i))\ ,
  \protect\label{E_MF}
\ee
where with $\cd m_i$ we define the lattice sum over the $6$ first neighbor
variables.

If this equation admits only one solution there is no ambiguity. If,
on the contrary, there are many solutions, one has to weight
(according to the previous discussion) different solutions with a
weight proportional to the exponential of minus the free energy
(multiplied by $\beta$).

Our ideal goal is to look for all solutions of this equation,
which correspond to local minima of the free
energy, but this is an awful task when the number of solutions
is very large, as it happens at low $T$.  Here we have just looked for
the solutions with higher, positive and negative, magnetization,
($m_+$ and $m_-$) using a simple iterative scheme.  We have started
the iterative procedure used to solve eq. (\ref{E_MF}) from the two
initial conditions $m_i=m_s$ and $m_i=-m_s$. Although a completely
safe procedure would start
from $m_s=1$, it is more convenient (and it does not change the
results) to take a value for $m_s$ slightly smaller than one. The
appropriate value of $m_s$ depends on the temperature; in our
simulations we have taken $m_s=.6$.

In the high $T$ regime both these runs converge to the same
(unique) solution. In a broken phase they will tend to different
solutions with average magnetization of opposite signs. This procedure
should be good enough to localize the temperature $T$ below which the
solution of (\ref{E_MF}) is not unique, and to give relevant
quantitative hints about the structure of the phase transition.

We label the solutions of the mean field equations, in a given
realization of the magnetic field, by the index $\alpha$; given the
pattern of our search $\alpha$ is limited to take only one or two
values.  For each realization of the magnetic field the index $\alpha$
belongs to the set $A$ (which can be, in our simulation, constituted
of $1$ or $2$ solutions). The average over different field samples
(which we denote by a bar: we denote the average over different
solutions by $\langle \cdot \rangle$) is done by having $A$ running
from $1$ to $N_A$.

In each solution $\alpha$ (characterized by the $V\equiv L^3$ values of the
local magnetization $m_i$) we compute the relevant observables. We
define the total magnetization density

\be
  m^\alpha \equiv \usv \sum_{i} m^\alpha_i \ ,
\ee

and the sum of the squared local variables

\be
  q^\alpha \equiv \usv \sum_{i} (m^\alpha_i)^{2} \ .
\ee

We define the energy density

\be
  E^\alpha \equiv - \usv \sum_i
  ( \frac{1}{2} m^\alpha_i \cd m^\alpha_i + h_i m^\alpha_i )\ ,
\ee

the entropy density

\be
  S^\alpha \equiv - \usv \sum_i
  ( \pap \log ( \pap )  + \pbp \log ( \pbp ) ) \ ,
\ee

and the total free energy as

\be
  F^\alpha = V (\beta E^\alpha - S^\alpha) \ .
\ee

The weight $w_\alpha$ associated to each solution $\alpha$ is given by

\be
  w_\alpha= \frac{\exp(-\beta F^\alpha)}{Z_w}\ .
\ee

\section{Numerical Results for Local Quantities}

Here we present numerical results for system of size up to $48$ in a
range of $\beta$ that goes from $1.1$ to $1.5$ (we will always give
$\beta$ in units of the critical $\beta$ at zero random field, i.e.,
${1 \over 6}$). We have analyzed $600$ random field
samples for the $12^3$ lattice, $400$ for the $24^3$ lattice, $200$ for
the $36^3$ lattice and $30$ for the $48^3$ lattice.

In this section we will discuss the behavior of local quantities (i.e.,
those objects that can easily be constructed from the magnetization),
while in the next section we will concentrate our attention on the
response functions, which must be computed by inverting the lattice
equivalent of $(-\Delta+V''(\phi))$, a highly non-local operation.

A very interesting quantity is

\be
   W^2 \equiv \sum_\alpha w_\alpha^2\ .
\ee
This quantity is different from $1$ when the mean field equations admit
more than one solution: roughly speaking $W^{-2}$ is the
average number of relevant solutions.  We display the results for
$W^2$ as function of $\beta$ in fig.~1. We see that $W^2$ becomes
sizably different from $1$ only at $\beta$ greater than $1.2$. We see
a change in regime at this beta, which we denote by $\beta_1$.

Another quantity that is interesting to measure is the maximal
magnetization $m_M^2$, defined as $\max_\alpha (m^{\alpha})^2$. In
fig.~2 we show the $\beta$ dependence of $m_M^2$ for different lattice
sizes.  We see a transition from an asymptotic zero value of $m_M^2$
to a non zero value around $\beta=1.35$. The transition becomes
sharper by increasing the size of the lattice. We see a change in
regime also at this new value of $\beta$, which we denote $\beta_2$.

A more detailed understanding can be obtained by considering the
correlation functions of the local magnetization. At this end we define, for
each solution $\alpha$, the magnetization on a $2$-plane as

\be
  M^\alpha_x(\lambda) \equiv \sum_{y,z} m^\alpha(\lambda,y,z)\ .
\ee

$M^\alpha_y(\lambda)$ and $M^\alpha_z(\lambda)$ are defined in an
analogous way. We define the zero (bi--)momentum
magnetization-magnetization correlation function
for the solution $\alpha$ as

\be
  C^\alpha(\lambda) \equiv
  \sum_{\mu=x,y,z; \ \lambda_1, \lambda_2 \ \mbox{\tiny such that}\
  |\lambda_1-\lambda_2|= \lambda}
  M^\alpha_\mu(\lambda_1) M^\alpha_\mu(\lambda_2)
  \ .
\ee

The total correlation function at distance $\lambda$, averaged over
$N_A$ samples, is defined as

\be
  C(\lambda) \equiv \frac{1}{N_A} \sum_A \sum_\alpha
  w^\alpha C^\alpha(\lambda)\ ,
\ee
and we denote by $C_c(\lambda)$ its connected part.

At first order in perturbation theory in the strength of the random
field \cite{PARISA,PARSOU,YOUNG} $C_c(\lambda)$ has (as we have discussed
before) a double pole in Fourier space. It has also been shown that
this form retains its validity at all orders in perturbation theory,
and should be exact in the region where supersymmetric considerations
hold.  In $x$ space that leads to

\be
  C_c(\lambda) \simeq A (1+\frac{\lambda}{\xi^{(C)}})
  e^{-\frac{\lambda}{\xi^{(C)}}} + B\ ,
  \protect\label{F_COR}
\ee
that defines the correlation length $\xi^{(C)}$.

In fig.~3 we plot the inverse correlation length as a function of
$\beta$. We have used a global fit to $C(\lambda)$ (which has
determined $\xi^{(C)}$, $A$ and $B$, by assuming a functional
dependence that takes in account the periodic boundary conditions).
In all cases we have computed the statistical errors by using a
standard jack--knife procedure. We have also computed $\lambda$
dependent correlation length estimators. By averaging them in the
plateau region we have obtained another estimate of $\xi^{(C)}$, which
turns out to be completely compatible with the one coming from the
global fits.  The fits turn out to be of very good quality, confirming
the approximate validity of the form (\ref{F_COR}).

The correlation length of fig.~3 has quite a broad maximum
close to $\beta=1.35$. $\xi^{(C)}$ close to its peak increases when going
from $L=12$ to $L=24$, but for larger lattices it remains constant.

In fig.~4 we plot the maximum value of the correlation length,
$\xi^{(C)}_m$, as a function of $\frac{1}{L}$, to stress the
saturation that occurs for large $L$. The asymptotic $\xi^{(C)}_m$ is
of order $4.5$.  It is rather consistent that the correlation length
becomes independent from the size for sizes $3$ to $4$ times larger than the
correlation length.

In fig.~5 we plot the coefficient $B$ (i.e., the constant asymptotic
value of the correlation function) computed from the fit to
$C(\lambda)$ as a function of $\beta$. For large volumes $B$ should
become identical with $m^2$ (which, in our analysis, turns out to be
very similar to $m_M^2$), but its finite size corrections are smaller,
especially in the high temperature region, where $B$ and $m_M^2$ are
asymptotically zero.  $B$ seems to take a non-zero expectation value
starting from $\beta_2 \simeq 1.35$.  This method gives a very good
estimate of the value of the critical temperature where $m_M^2$
becomes sizably different from zero.

In the region where $m_M^2$ is zero all different
solutions of the mean field equations should become locally equal in
the infinite volume limit, or more precisely their absolute difference
should be in average go to zero with the volume.

It is natural to ask if these values of $\beta$ do correspond in the
thermodynamical limit to real phase transitions.  The quantity $W^2$
becomes different from zero as soon as there exist a realization of
the magnetic field that admits two solutions.  An explicit
computation shows that if $h(i)=(-1)^{x+y+z}$ one finds two solutions
when $\beta \ge \beta_G \simeq 1.015$\footnote{An approximate
formula valid for small $\ch$ is $\beta_G= 1+{\ch^2 \over 144}+O(\ch^4)$.}.
Simple minded arguments (which generalize the original Griffiths
theorem for random diluted magnetic systems) suggest that the free
energy is $C^\infty$ but not analytic at $\beta_G$

For $\beta > \beta_G$ there exist realizations of the magnetic field
for which at least two solutions exist.
These field configurations are special, and their measure is
small. We expect therefore
that $W^2$ is different from $1$, mathematically speaking, for
$\beta\ge\beta_G$, but it becomes sizably different from $1$ only at
$\beta \ge \beta_1$. Similar arguments can be done for $\beta_2$. The
non vanishing of $m_M^2$ for $\beta \ge \beta_G$ is a pathology that
arises from our choice of considering only the maximal solution. If we
consider the physically relevant quantity, i.e.,

\be
  m^2=\sum_\alpha w_\alpha \ (m^\alpha)^2\ ,
\ee
it should become different from zero only at values of $\beta$ much
higher than $\beta_G$. The fact that the correlation length remains
finite and somewhat small near $\beta_2$ may be taken as an
indication that the true ferromagnetic transition at which $m^2$
becomes different from zero is at higher values of $\beta$.

The situation would be clarified if we could compute the full expression for
$C(x)$, summing over all the solutions, but we have left this task for
a future work.

\section{Numerical Results for the Response Functions}

To compute the correlation functions in the mean field
approach we must use the fluctuation-dissipation theorem. We are
therefore lead to consider the susceptibility function $\chi_{i,j}$,
which is equal to the derivative of the magnetization $m_i$ with
respect to the field $h_j$ (for sake of typographical clearness in the
following we will omit the solution label $\alpha$). If there is a
single stable state we have to perturb the unique solution of equation
(\ref{E_MF}). We get in this way the equation

\be
 \chi_{i,j} = \beta (1-m_i^2)(\cd \chi_{i,j} + \delta_{i,j}) \ .
\ee
This is a linear sparse equation that can be solved by using standard
iterative techniques.

The computation of $\chi$ for all the value of $i$ and $j$ would be
extremely time consuming, so we compute the Green functions $g_i
\equiv \chi_{i,0}$ by setting $j=0$ and iterating the relation

\be
  g_i = \beta (1-m_i^2)(\cd g_i+\delta_{i,0}) \ .
\ee

We also compute the susceptibility
$\chi \equiv \frac{1}{V}\sum_{i,j}\chi_{i,j}$
by iterating

\begin{eqnarray}
  \nonumber \chi_i & = & \beta (1-m_i^2)(\cd  \chi_i+ 1) \ ,\\
  \chi & \equiv & \frac{1}{V}\sum_{i=1}^V \chi_i\ ,
\end{eqnarray}

and the {\em overlap} susceptibility
$\chi^{q} \equiv \frac{1}{V}\sum_{i,j}\chi_{i,j}m_i m_j$ from

\begin{eqnarray}
  \nonumber \chi^{q}_i & = &
  \beta (1-m_i^2)(\cd  \chi^{q}_i+ m_i)\ , \\
  \chi^{q} & \equiv & \frac{1}{V}\sum_{i=1}^V \chi^{q}_i m_i\ .
\end{eqnarray}

The name {\em overlap susceptibility} arises from the following
considerations. Let us consider two replicas ($\sigma$ and $\tau$) of
the same system whose dynamics is determined by a Hamiltonian that
contains the usual one system contribution plus a direct coupling
among the two systems. The total Hamiltonian is

\be
  H[\sigma ] + H[\tau ]- \epsilon \sum_i \sigma_i \tau_i \ .
\ee

This construction is common in the study of other disordered
systems like spin glasses. The quantity $\chi^{q}$ coincide with
${\partial q \over \partial \epsilon}$, evaluated at $\epsilon=0$,
where $q$ is the overlap density, i.e., ${1 \over V} \sum_i \sigma_i
\tau_i$.

In the interesting case in which the mean field equations admit many
solutions $\alpha$ we follow the simplest procedure of weighting each
of these with the weight $w^\alpha$ (we remind that we are taking in
account only the two maximal solutions). In this way we are obtaining
only one term of the two that form the full susceptibility.
It is easy to check that the response function

\be
  R(i,j) \equiv {\partial \over \partial h_j}
  \sum_\alpha w_\alpha m^\alpha_i
\ee
is given by

\be
  R(i,j) = \chi_{i,j} + \beta \big (  \sum_\alpha w_\alpha m_i^\alpha
m_j^\alpha
                              - \sum_\alpha w_\alpha m_i^\alpha
                                \sum_\gamma w_\gamma m_j^\gamma\ \big)\ .
\ee
The second term (in brackets),
which arises in presence of many solutions, is likely
to be dominant near the critical point, as will shall see below. It may
be convenient to call the first the diagonal contribution, and the
second one the off-diagonal contribution.

We have computed the diagonal contributions $\chi$ and $\chi^{q}$ with
the results shown in figs.~6 and 7.  It is impressive that $\chi$ has
a sharp maximum close to $\beta_1$, while $\chi^{q}$ has a peak at
much higher beta (slightly above $\beta_2$) and does not show any
significant anomaly at $\beta_1$. These two peaks are volume
independent for large volume.  The correctness of this result is
confirmed by the direct analysis of the correlation length
corresponding to $\chi$, $\xi_\chi^{(1)}$, which we show in fig.~8.
$\xi_\chi^{(1)}$ does never become large in the whole region and for
$\beta \le \beta_1$ essentially coincides with $\xi^{(C)}$, the
correlation length we have discussed in the previous section. We find that
the supersymmetry prediction of equality of the two correlation lengths is
correct in the region $\beta \le \beta_1$ where only one solution is
present.

We have also considered the correlation lengths $\xi_\chi^{(n)}$
defined by taking the $n$-th power of the zero bi-momentum correlation
functions that give $\xi_\chi^{(1)}$, and by looking at their decay.
They do not present a significant difference (once divided by $n$) from
the one obtained for $n=1$. In fig.~9 we show the
correlation length with $n=2$ (which has the smallest statistical error),
which can be compared with the $n=1$ result of fig.~8. The two sets of curves
are very similar.

Evaluating at least some approximation to the off-diagonal
contribution to the susceptibility is essential. We have done it
by only using our maximal solutions.
There is a large
statistical error. In the low temperature region we expect that
the off-diagonal contribution is proportional to $N^{1/2}$, this contribution
arising from a few exceptional configurations of the magnetic field
that have two solutions with opposite magnetization with similar
weight. This event happens with a probability of order $1/N^{1/2}$; the
corresponding off-diagonal susceptibility is of order $N$, so that the net
contribution to the susceptibility coming from these exceptional
configurations is proportional to $N^{1/2}$. In this region the off-diagonal
susceptibility is dramatically increasing, showing the trend to diverge
about $\beta \simeq 1.30$. Anyhow there is no convincing argument
that implies that the restriction to the maximal solution should be a good
approximation, apart from very close to $\beta_1$, where only two stable
solutions are expected.

A full computation (including all the solutions) of both the diagonal and the
non-diagonal contribution to the susceptibility would be extremely
interesting.

\section{Conclusions}

The existence of many solutions to the mean field equations turns out
to be a crucial phenomenon; inside a single solution (at least of the
maximal type) one does not see any sign of the presence of a divergent
correlation length. The critical behavior of the susceptibility and of
the correlation length in a $3d$ RFIM is dominated by the effects of
the presence of many solutions. The supersymmetric predictions start
to fail exactly at the point where one finds more than one solution of the mean
field equations. It is not surprising that dimensional
reduction, which completely misses the existence of more than one
solution, gives unreliable exponents at the critical point.

It is reasonable that each solution of the mean field equation does
correspond to a valley for the energy in configuration
space\footnote{We reserve the word state for solutions (or cluster of
solutions) such that their distance ${1 \over V} \sum_i (m_i^\alpha -
m_i^\gamma)^2$ does not vanish in the infinite volume limit.}.
In this case the dynamics of Monte Carlo simulations of a real
system also at temperature slightly above the critical one is likely
to be dominated by thermal activated tunnelling among different
valleys, and we expect it to be a slow process.

\section*{Acknowledgements}
We thank M. Aizemann, M. Mezard and N. Sourlas for many interesting
discussions.

\vfill
\newpage

\section*{Figure Captions}

\begin{enumerate}
\item
  \protect\label{F_W2}
  $W^2$ as a function of $\beta$. dots for the $12^3$ lattice, dashes-dots
  for the $24^3$ lattice, dashes for the $36^3$ lattice and solid line
  for the $48^3$ lattice.
\item
  \protect\label{F_MM2}
  As in fig.~1, but $m_M^2$.
\item
  \protect\label{F_CSIB}
  The inverse correlation length $m\equiv\frac{1}{\xi^{(C)}}$
  as a function of $\beta$ for different lattice sizes.
\item
  \protect\label{F_CSIL}
  The maximum correlation length $\xi^{(C)}_m$ as a function of the inverse
  linear size of the system.
\item
  \protect\label{F_CONST}
  The constant coefficient $B$ from the global fit to $C(\lambda)$ as
  a function of $\beta$.
\item
  \protect\label{F_CHI}
  As in fig.~1, but $\chi$, the diagonal contribution to the susceptibility.
\item
  \protect\label{F_CHIQ}
  As in fig.~1, but $\chi^q$, the diagonal contribution to the overlap
  susceptibility.
\item
  \protect\label{F_CSI1}
  As in fig.~1, but $\xi^{(1)}_\chi$.
\item
  \protect\label{F_CSI2}
  As in fig.~1, but $\xi^{(2)}_\chi$.
\end{enumerate}
\vfill
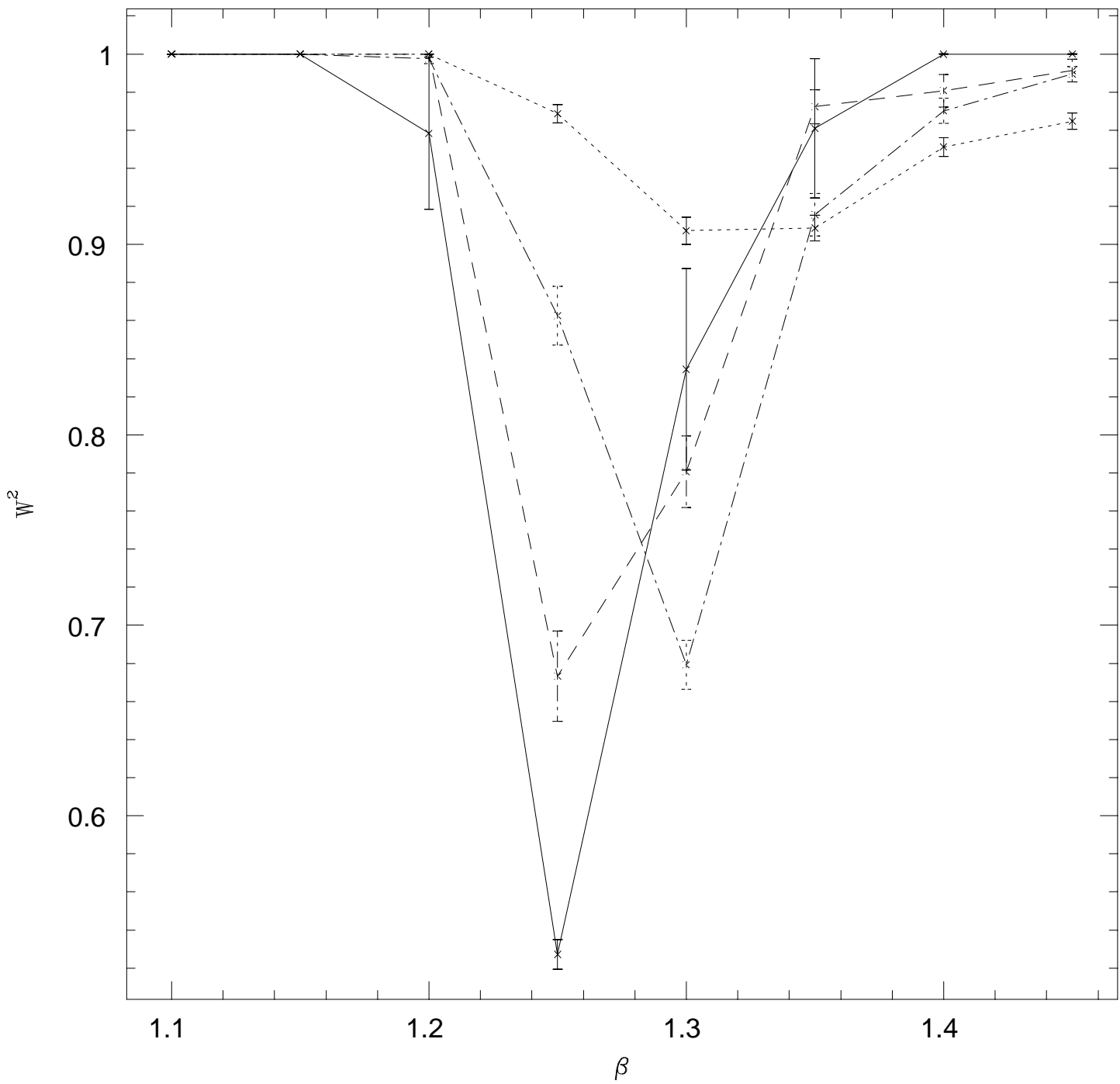
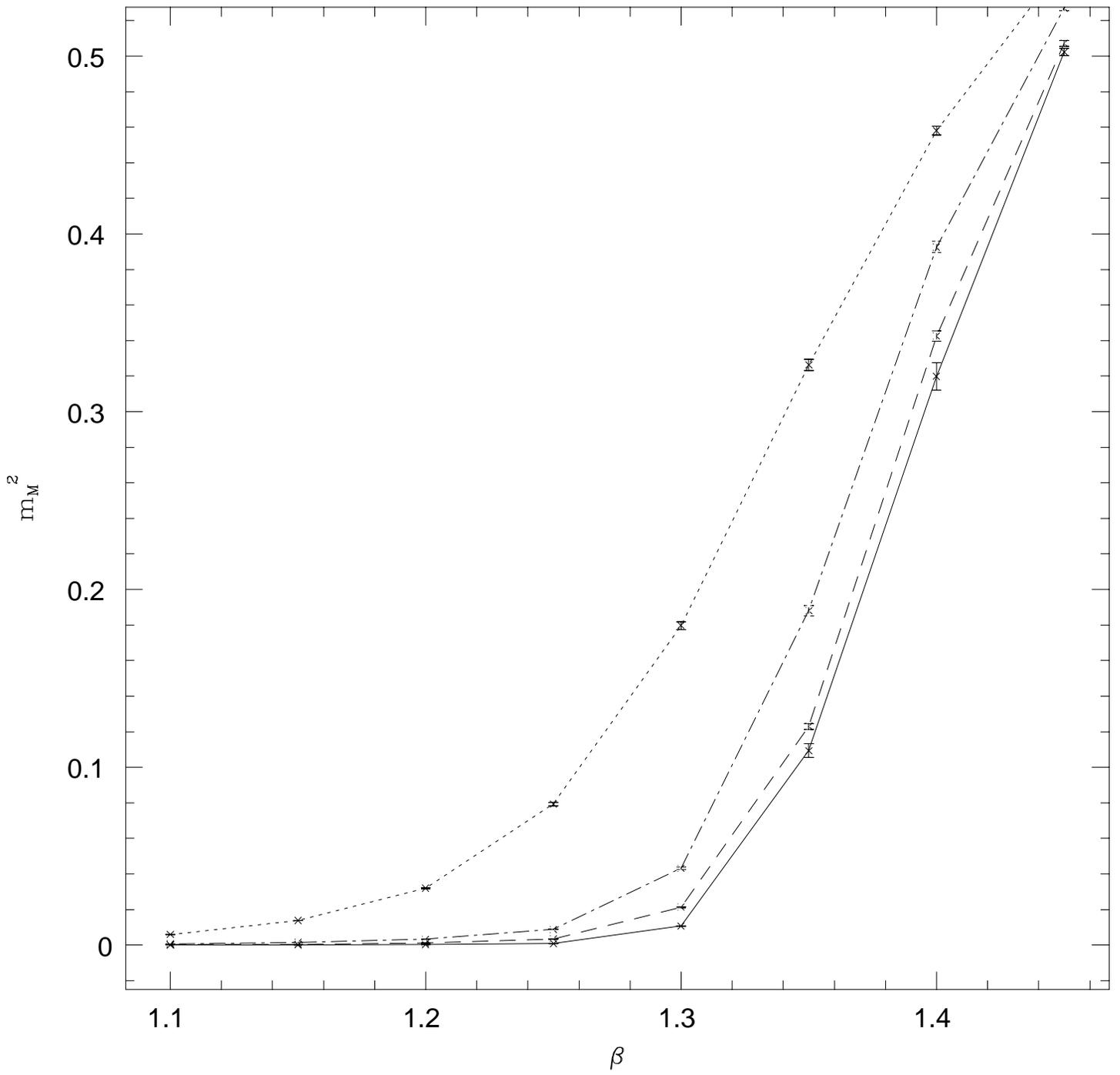
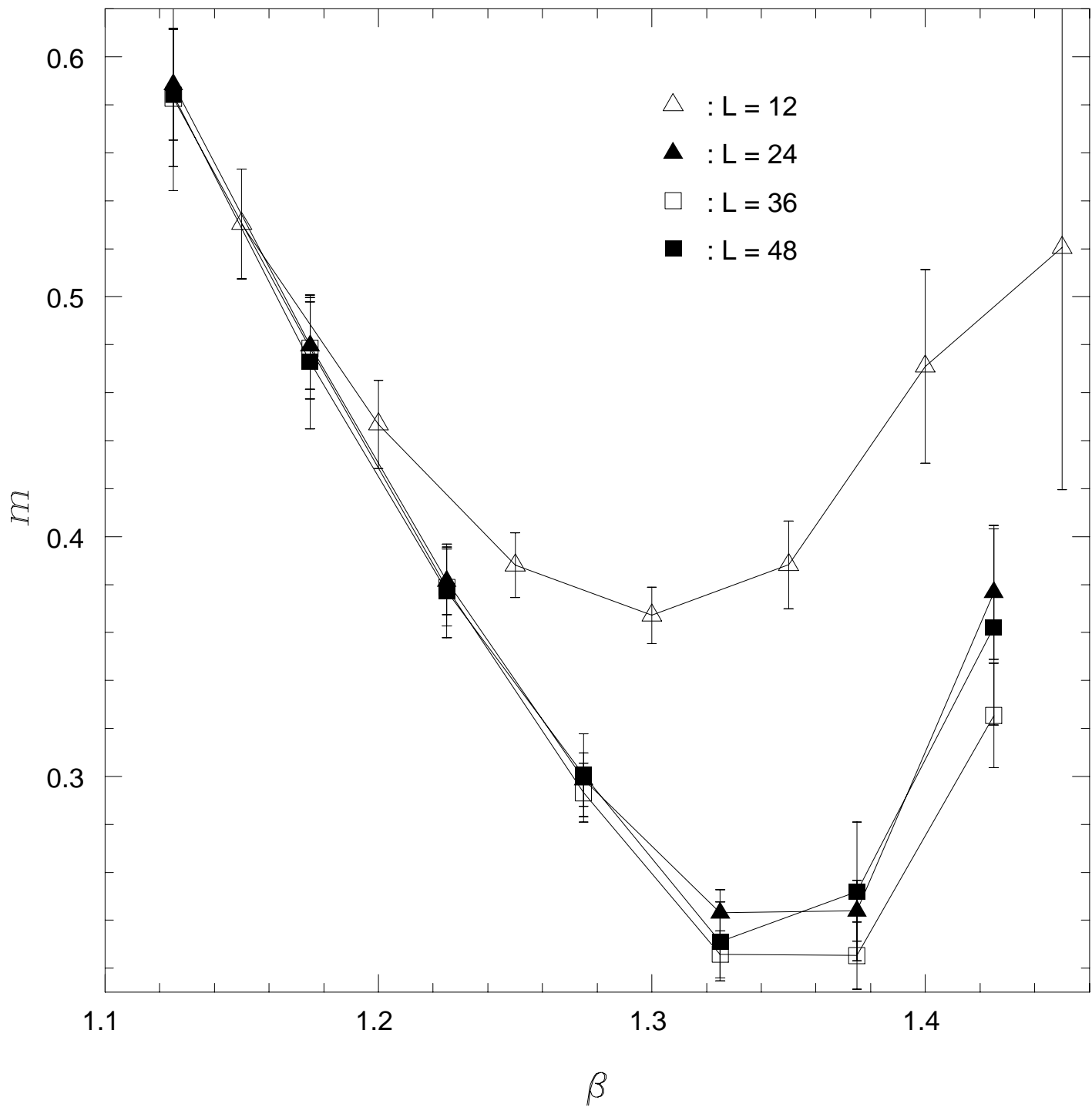
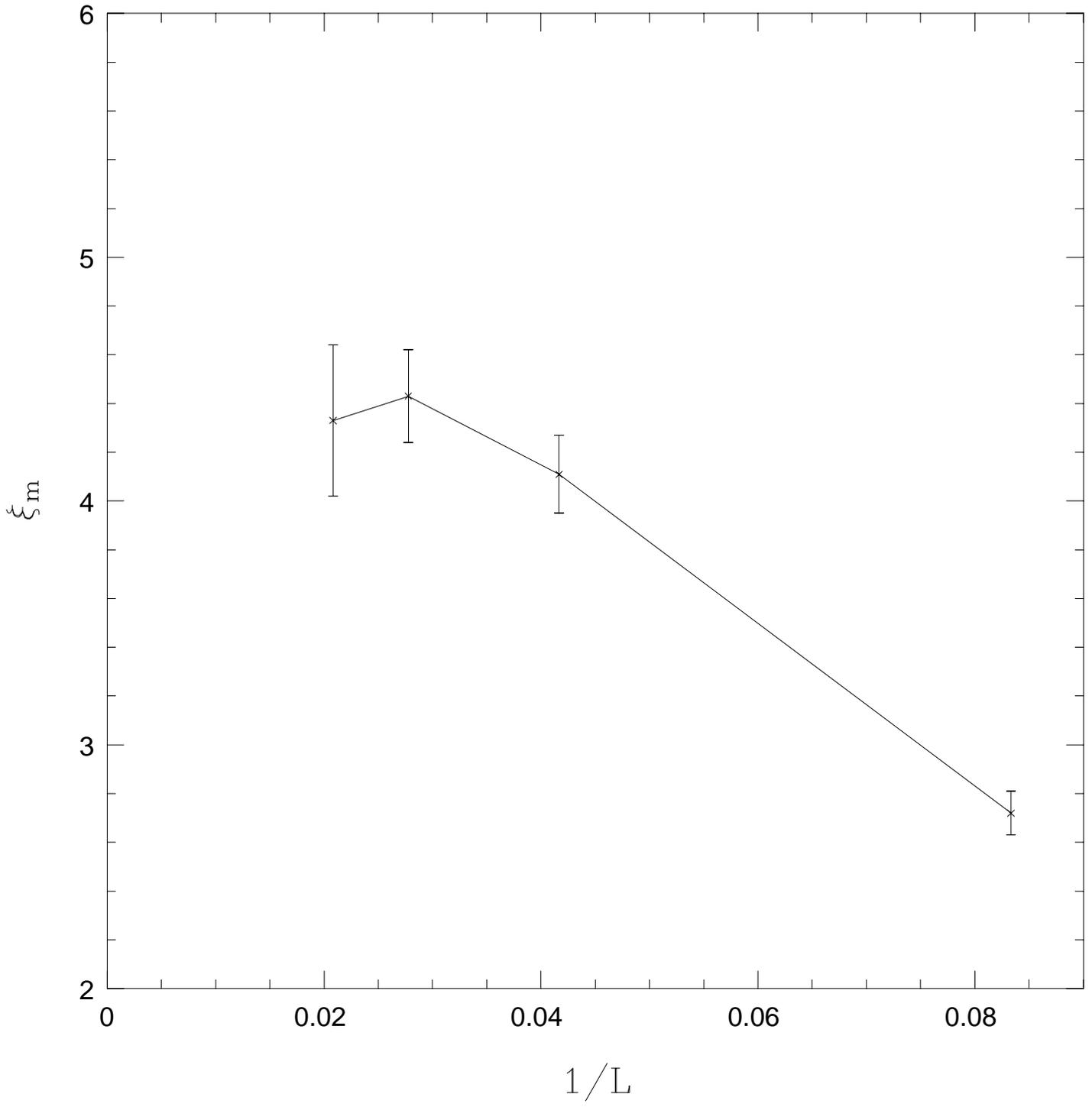
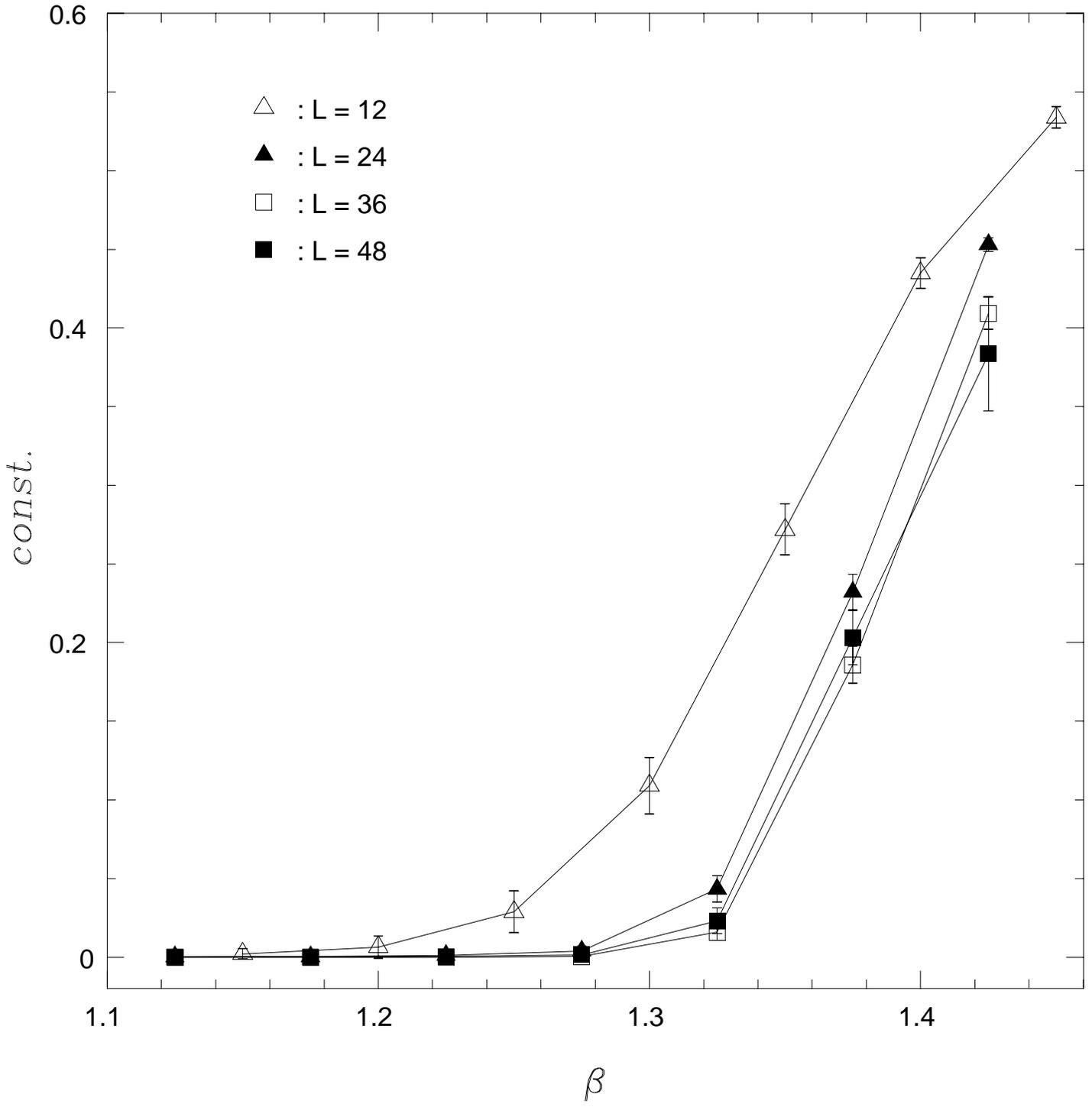
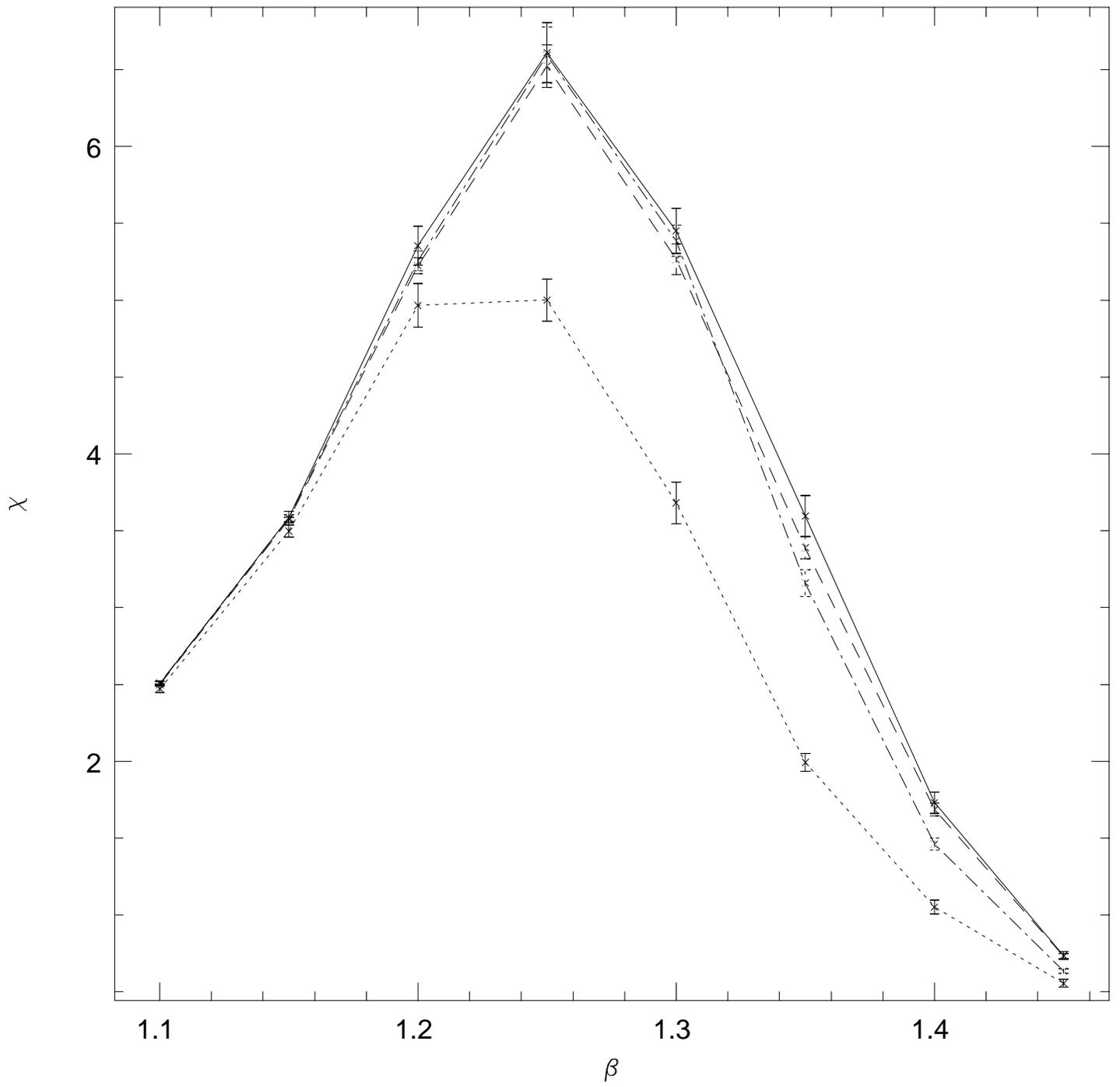
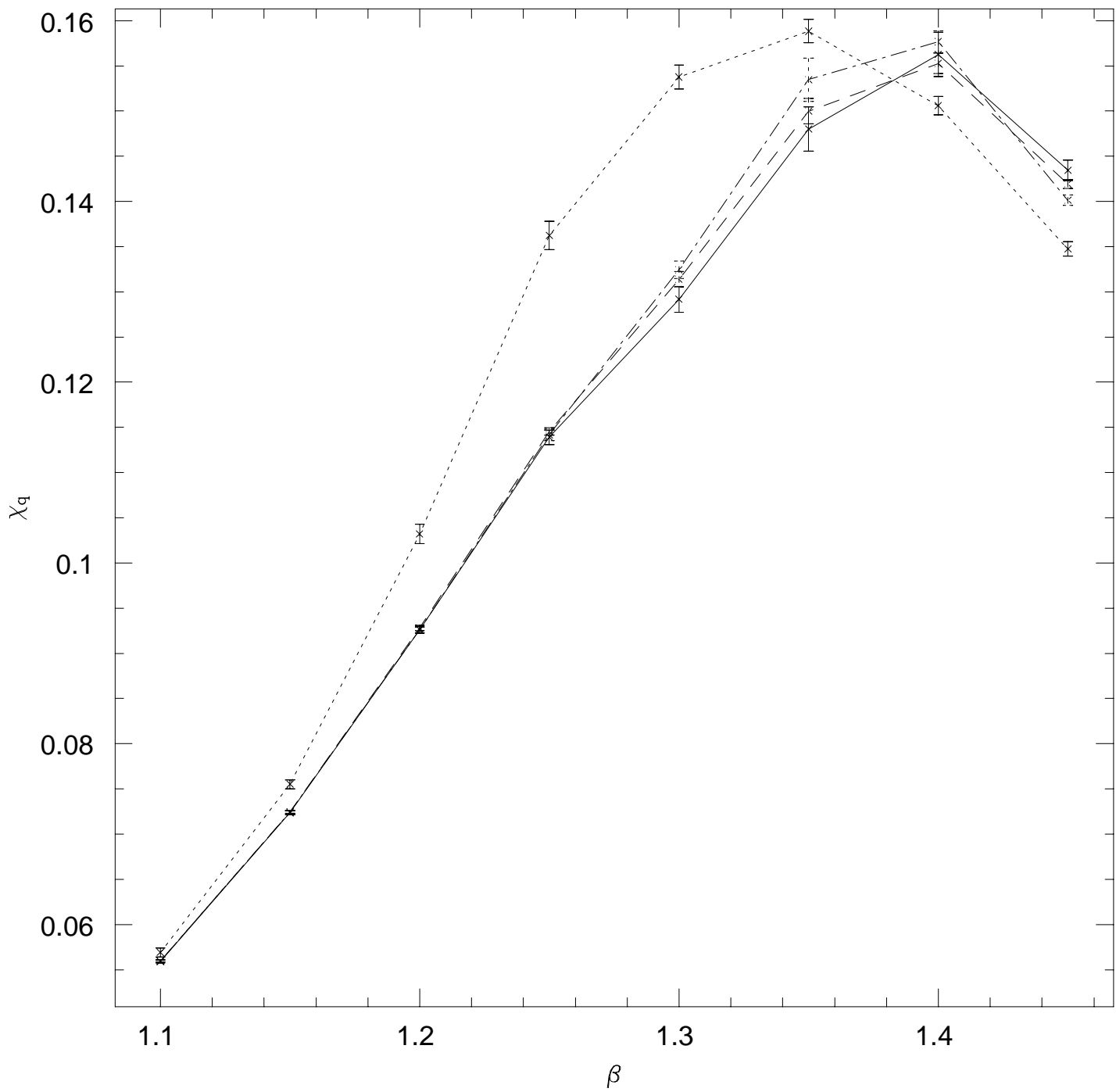
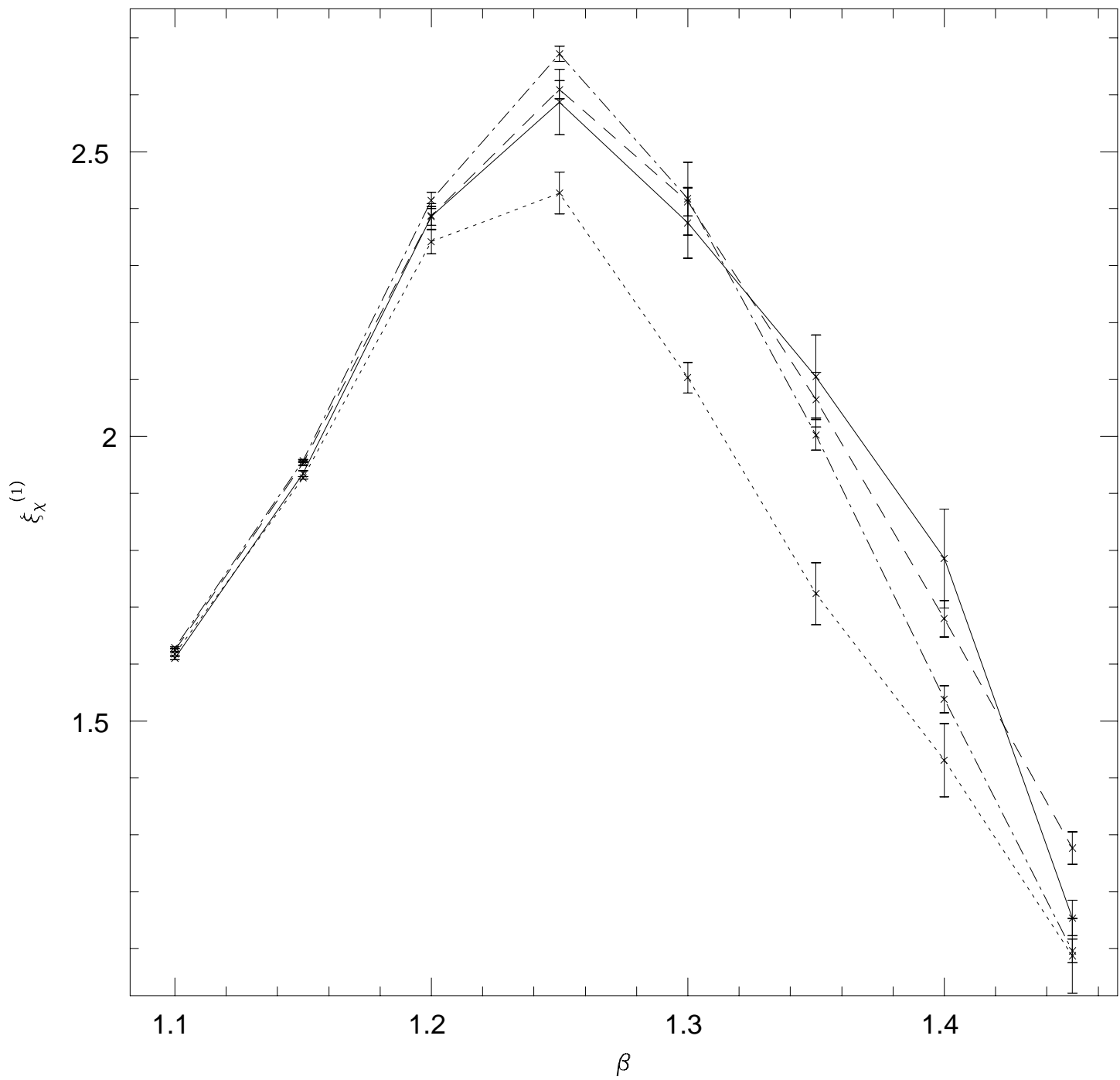
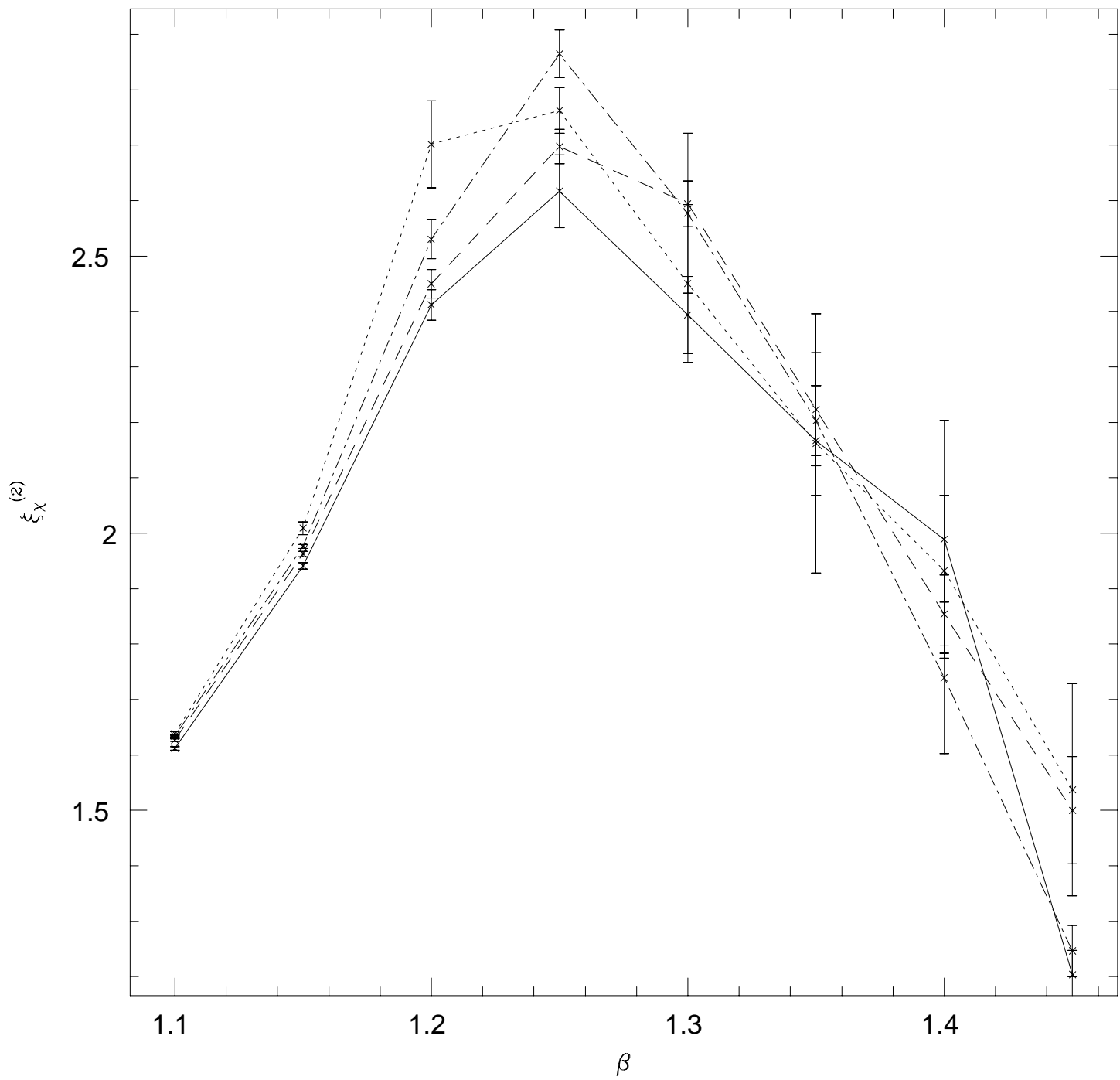
\end{document}